\documentclass[%
reprint,
showpacs,preprintnumbers,
amsmath,amssymb,
aps,
pre,
]{revtex4-1}

\usepackage{graphicx}% Include figure files
\usepackage{dcolumn}% Align table columns on decimal point
\usepackage{bm}% bold math
\usepackage{braket}
\usepackage{color}
\usepackage{caption}
\usepackage{ragged2e}
\usepackage{natbib}% add hypertext capabilities
\usepackage{hyperref}
\usepackage{url}
\usepackage[utf8]{inputenc}

\DeclareCaptionFormat{myformat}{#1#2(Color online) #3}
\DeclareCaptionJustification{justified}{\justifying} 
\newenvironment{myfigure}%
{\captionsetup{format=myformat,labelsep=period,justification=justified, font=small}
\figure
}%
{\endfigure}%

\begin{document}
\preprint{APS/123-QED}
\title{State Transitions and Decoherence in the Avian Compass}% Force line breaks with \\
\author{Vishvendra Singh Poonia}
\email{vishvendra@iitb.ac.in}
\author{Dipankar Saha}
\author{Swaroop Ganguly}%
 \email{sganguly@ee.iitb.ac.in}
\affiliation{%
 Department of Electrical Engineering, Indian Institute of Technology Bombay, Mumbai -- 400076, India\\
 }%

\date{\today}%

\begin{abstract}
The radical pair model has been successful in explaining behavioral characteristics of the geomagnetic
compass believed to underlie the navigation capability of certain avian species. In this study, the spin 
dynamics of the radical pair model and decoherence therein are interpreted from a microscopic state 
transition point of view. This helps to elucidate the interplay between the hyperfine and Zeeman 
interactions that enables the avian compass, and the distinctive effects of nuclear and environmental 
decoherence on it. Using a quantum information theoretic quantifier of coherence, we find that 
nuclear decoherence induces new structure in the spin dynamics without materially affecting the 
compass action; environmental decoherence, on the other hand, completely disrupts it.
\end{abstract}
\pacs{Valid PACS appear here}% PACS, the Physics and Astronomy
\maketitle
\section{Introduction}
\label{Intro}
Overt quantum effects seem to play a key role in the functionality of a number of biological systems, e.g. excitonic transport in 
photosynthetic pigments, coherent spin dynamics in avian magnetoreception, inelastic electron tunneling in olfaction, and hydrogen tunneling in  
enzyme  catalysis~\citep{engel2007evidence, plenio2008dephasing, lambert2013quantum, mohseni2008environment, 
scholes2011lessons, turin1996spectroscopic, brookes2007could, nagel2006tunneling, wiltschko1972magnetic, schulten1978biomagnetic, wiltschko2010directional}.
This is not only intriguing from a physics perspective – given the noisy high-temperature ambience in these 
situations – but also promises to reveal robust ways to harness ‘quantumness’ for engineering new and improved systems, both biomimetic and otherwise.

In this work we study avian magnetoreception, which is responsible for the geomagnetic field assisted navigation ability of various bird 
species~\citep{wiltschko1972magnetic, schulten1978biomagnetic}. A radical pair (RP) model, comprising photo-excited unpaired spins along with an 
anisotropic hyperfine interaction, has been proposed as a possible mechanism~\citep{steiner1989magnetic, ritz2000model,solov2009magnetoreception,solov2012reaction}. 
This kind of a model is supported both by spin chemistry findings~\citep{timmel2004study, miura2006spin, rodgers2009magnetic, lau2014spin} and by behavioral experiments 
on certain bird species, e.g. the European Robin~\citep{ritz2000model, wiltschko2010directional, wiltschko1993red, ritz2009magnetic}.

One of the central quests in RP model studies is the determination of the role of quantum effects especially coherence, including nuclear and environmental 
decoherence effects, in the spin dynamics~\citep{tiersch2012decoherence, ritz2004resonance, cai2010quantum, cai2013chemical, kominis2009quantum}. A few 
groups~\citep{gauger2011sustained, bandyopadhyay2012quantum, ritz2004resonance, ritz2009magnetic} have established the presence of long coherence in RP spin states. 
Others~\citep{cai2013chemical}  have quantified coherence using a quantum interferometer analogy and statistically concluded that global electron-nuclear 
coherence is a resource for chemical compass by observing sensitivity as a function  of global coherence. Tiersch and Briegel identify 
nuclear decoherence as a necessary ingredient for the magnetosensitive spin dynamics of the RP system~\citep{tiersch2012decoherence}.  
Apart from coherence, Gauger et al. and Cai et al. have studied the role of entanglement in the compass action of the RP 
model~\citep{gauger2011sustained, cai2010quantum}. However, the distinct operational role of nuclear and environmental decoherence in RP spin dynamics is 
still unclear. Understanding this is essential to the appropriate selection/engineering of materials for solid state emulation  of RP spin dynamics, 
and possibly other quantum biomimetic applications.

In this work, we take a microscopic view of radical pair spin dynamics, analyzing the distinctive role of nuclear and 
environmental decoherence, and examine their specific effects in its magnetosensitive behavior. We look at the state 
transitions involved in  radical pair spin state evolution and elucidate the effect of nuclear  and environmental  
decoherence  on these transitions. Our conclusions are validated by applying an information theoretic measure of coherence.  
Further, our spin transition point of view provides new insights into the role of Zeeman and hyperfine interactions in the
magnetosensitive dynamics of the RP spin system. We also revisit some of earlier RP model results from this  new perspective.

The salient characteristics of avian magnetoreception have been demonstrated by multiple behavioral experiments. Firstly, it exhibits a certain dynamic 
range around the geomagnetic field. This dynamic range behavior is versatile, in that it adapts to a new Zeeman field if the bird is exposed long enough
to it~\citep{ritz2009magnetic}. Secondly, the compass action is found to be disrupted by an external  RF field of a particular frequency. 
Both of these features are, in fact, well-explained within the RP model~\citep{ritz2004resonance,ritz2009magnetic, Poonia2014FunctionalWindow}.

In the RP system, we effectively have a three spin system evolving under a Hamiltonian that contains two interactions -- hyperfine and Zeeman.  
The singlet and triplet radicals recombine distinctly thus leading to different products. The final yield corresponding to singlet and 
triplet states depends on the magnetic field. Strikingly, however, we find that the Zeeman interaction alone is not 
sufficient to make the final yields dependent on the magnetic field orientation (angle between geomagnetic field and radical pair axis); neither can the hyperfine 
interaction on its own cause spin transitions from the singlet to all three triplet states, and thereby impart magnetic sensitivity. It is
the interplay between the Zeeman and the hyperfine interactions that makes the overall spin dynamics magnetosensitive.

This paper is organized as follows: In Sec.~\ref{SpinDynamicsRP}, we investigate the state transitions involved in the 
RP spin dynamics. In Sec.~\ref{NuclearDecoherence} and~\ref{EnvironDecoherence}, we examine the 
roles of nuclear and environmental decoherence respectively. In Sec.~\ref{Conclusion}, we present our 
conclusions and perspective for solid state emulation of the avian compass. 
\section{Spin dynamics of the radical pair}
\label{SpinDynamicsRP}
In  order  to  study the spin transitions involved in the RP dynamics, we choose a representative RP system in which an unpaired 
spin on each of two radicals, and a single  nucleus on one of them is responsible  for  an hyperfine  interaction therein. This model may 
be directly extrapolated to multinuclear systems~\citep{gauger2011sustained, cai2012quantum}. The nucleus preferentially interacts with the spin on the same radical and both 
the spins interact with the geomagnetic Zeeman field. Therefore, the RP Hamiltonian looks like~\citep{gauger2011sustained}:
\begin{eqnarray}
\label{eq1}
H =\gamma \mathbf{B} \cdot (\hat{S_1} + \hat{S_2}) + \hat{I} \cdot \mathbf{A} \cdot\hat{S_2}
\end{eqnarray}
$\hat{S_1}$ and $\hat{S_2}$ are electron spin operators, and $\hat{I}$ is the nuclear spin operator given 
as: $\hat{I}, \hat{S_1}, \hat{S_2} \in \frac{1}{2}(\sigma_x, \sigma_y, \sigma_z)$, $\gamma = \mu_0 g$ 
is gyromagnetic ratio, $\mu_0$ is Bohr magneton and $g$ is electron g-factor (= 2).
We consider the illustrative case in which the hyperfine tensor $ \mathbf{A}=diag(0,0,a) $~\citep{cai2012quantum}. The external field (geomagnetic field) 
is characterized by $\mathbf{B}= B_0(sin\theta cos\phi, sin\theta sin\phi, cos\theta)$; $B_0 (= 47 \mu T)$ is the local geomagnetic field at 
Frankfurt~\citep{ritz2009magnetic} and $\theta$ is the magnetic field orientation. The axial symmetry of hyperfine tensor allows us to 
take $\phi=0$~\citep{gauger2011sustained}.
\begin{myfigure}[b]
\includegraphics[width=8.5cm,height=8.5cm,keepaspectratio]{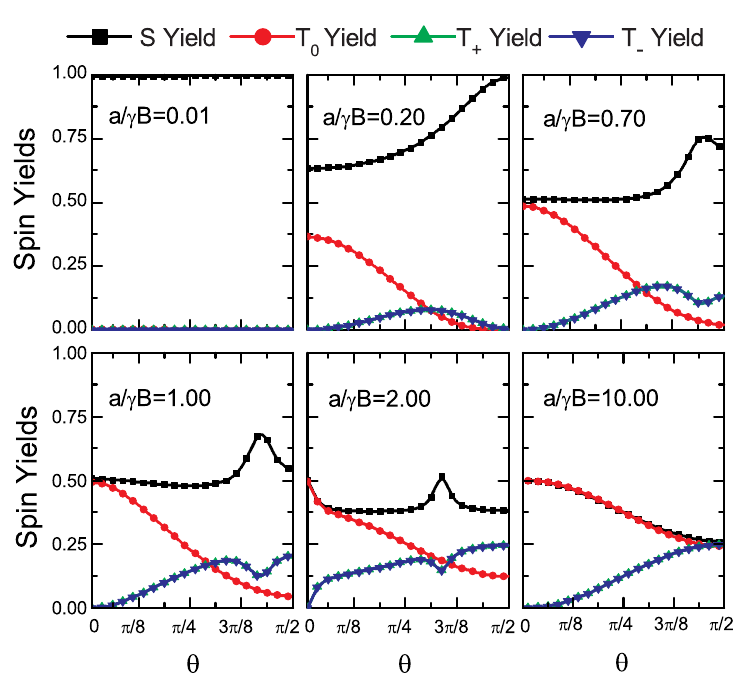}
\caption{Singlet ($S$) and triplet ($T_0$, $T_+$, $T_-$) yields vs. geomagnetic field orientation for various hyperfine coupling strengths. 
These figures illustrates how various spin transition pathways between $\ket{s}$, $\ket{t_0}$, $\ket{t_+}$, $\ket{t_-}$  
make the final spin yields angle dependent. Simultaneous Zeeman and anisotropic hyperfine plays collaborative role in inducing these 
transitions which ultimately lead to magnetosensitive product yield.(The various transitions induced by anisotropic hyperfine and Zeeman 
interactions are shown in the state transition diagram, Fig.~\ref{spindiag}.) Additionally, we observe a conspicuous peak appearing in S 
yield for a range of hyperfine constants.This peak corresponds to the dip in $T_+$ and $T_-$ yields. We recognize that this peak appears 
due to nuclear decoherence.} 
\label{fig1}
\end{myfigure}
\begin{myfigure}[t]
\includegraphics[width=6cm,height=6cm,keepaspectratio]{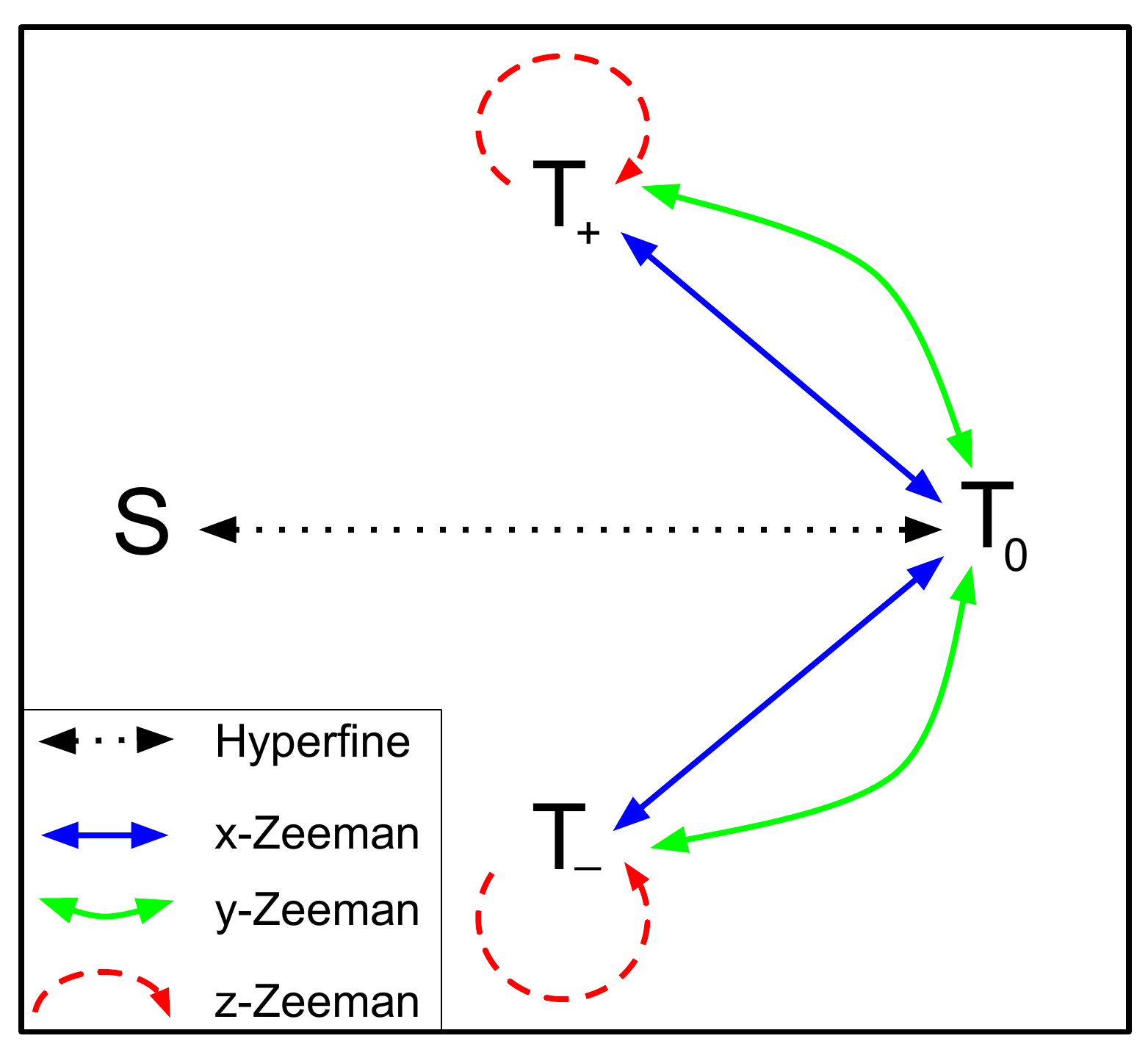} 
\caption{Spin-state transitions for various Hamiltonian interactions.} 
\label{spindiag}
\end{myfigure}
The RP spin dynamics starts at the instant of radical pair  generation, which is taken  to be  t = 0. The initial state of the radical pair is usually taken to be 
singlet state ($\ket{s}$) with the nuclear  spin completely depolarized  owing to  its interaction with  the neighboring soft  matter 
environment~\citep{ritz2000model,timmel1998effects,gauger2011sustained}.  We  adopt   the  quantum  master equation approach to simulate the dynamics of the RP system, 
similar to Gauger et al.~\citep{gauger2011sustained} but modified to highlight the exact spin transitions. Intrinsically,  the  Hilbert space  is eight 
dimensional. The  spin-dependent relaxation process happens through two channels -- the singlet  channel whereby radical  pairs in the $\ket{s}$ state recombine 
and triplet channel whereby radical pairs in the three triplet states $\ket{t_0}$, $\ket{t_-}$ and $\ket{t_+}$ recombine -- which are included as `shelving' states 
in the Hilbert space. In order to distinguish the spin transitions, we resolve the triplet channel into its three constituent channels corresponding  
to $\ket{t_0}$, $\ket{t_+}$ and $\ket{t_-}$. This is accommodated by augmenting the eight dimensional Hilbert space with four additional shelving states denoted 
as $\ket{S}$, $\ket{T_0}$, $\ket{T_+}$, and $\ket{T_-}$ .

This method is quite  versatile  for the calculation of the  yield corresponding  to  various  spin  states. The recombination of radical pair into singlet and 
triplet channels is modeled through decay operators in the master equation (ME) as:
$P_1 = \ket{S}\bra{s,\uparrow}$, 
$P_2 = \ket{S}\bra{s,\downarrow}$, $P_3 = \ket{T_0}\bra{t_0,\uparrow}$, $P_4 = \ket{T_0}\bra{t_0,\downarrow}$, $P_5 = \ket{T_+}\bra{t_+,\uparrow}$
, $P_6 = \ket{T_+}\bra{t_+,\downarrow}$, $P_7 = \ket{T_-}\bra{t_-,\uparrow}$ and $P_8 = \ket{T_-}\bra{t_-,\downarrow}$. 
The  Lindblad master  equation  describing  the evolution  of RP spin system is given by~\citep{gauger2011sustained}:
\begin{eqnarray}
\label{eq2}
\dot \rho = - \frac{i}{\hbar}[H, \rho] + k \sum\limits_{i=1}^8 P_i \rho P_i^\dagger - \frac{1}{2}(P_i^\dagger P_i \rho + \rho P_i^\dagger P_i)
\end{eqnarray}
Here \textit{k} ($=5 \times 10^5 s^{-1}$) is the singlet and triplet radical recombination rate. We note that this method is equivalent to the Haberkorn approach to modeling 
radical pair dynamics~\citep{gauger2011sustained, cai2013chemical,lau2014spin} but is more amenable for discerning the spin transitions involved in the compass action.
The system starts in the state $\rho(0) = \frac{1}{2}I \otimes (\ket{s} \otimes \bra{s})$. The ensuing spin evolution involves intersystem crossing 
between singlet and triplet states. It is accompanied by a spin dependent recombination process in which singlet and triplet radical  pairs  recombine through 
different channels; it is this Zeeman field dependent differential spin yield which is used by the avian neural system to sense the geomagnetic 
field~\citep{steiner1989magnetic, gauger2011sustained, cai2012quantum}, although the neurological processes involved  are not yet fully 
understood~\citep{bandyopadhyay2012quantum}. Our aim in this manuscript is to explain: one, the  details of the spin transitions responsible  for the magnetosensitive yield, 
and two, the  role  of coherent evolution  of electron  pair  spins and  decoherence  due to nucleus and environment in the overall functioning of chemical 
compass model of the avian magnetoreception.

In order to examine the spin transitions in the RP model, we simulate RP dynamics for a large number of hyperfine interaction strengths. For our choice of hyperfine 
coupling tensor, the Hamiltonian looks like:  
$\hat{H} = \gamma \mathbf{B} \cdot (\hat{S_1}+\hat{S_2})+a\hat{I_z}\hat{S_2^z}$.   
Here,  we vary  the  hyperfine coupling  strength from $\gamma B_0/100$ (very small compared  to  the Zeeman  strength) to  100$\gamma B_0$ (very large compared 
to the Zeeman strength) and  examine  the  singlet  and various triplet yields with respect to the orientation of the geomagnetic field. The results are presented 
in Fig.~\ref{fig1} from which we infer the following spin transitions are associated with various terms in the Hamiltonian: \\
(i) The hyperfine interaction induces the $\ket{s} \leftrightarrow \ket{t_0} $  transition but does nothing to $\ket{t_+}$ and $\ket{t_-}$ states. \\
(ii) The x-component of Zeeman  interaction induces the $\ket{t_-} \leftrightarrow \ket{t_0} \leftrightarrow \ket{t_+} $
transitions but leaves the $\ket{s}$ state alone. Similar is the case with y-component of Zeeman interaction.\\
(iii) The z-component of Zeeman interaction does not induce any inter-spin transition.\\
These results have been confirmed analytically in Appendix~\ref{Appen_SpinTransitionsRP}. They are also in agreement
with the findings of B. M. Xu et al. where these transitions follow from inspection of the Hamiltonian in 
\{$\ket{s}$, $\ket{t_0}$,$\ket{t_+}$, $\ket{t_-}$\} basis~\citep{xu2014effect}. 
The spin-state transitions induced by the hyperfine and Zeeman interactions as obtained above are summarized in 
Fig.~\ref{spindiag}.% Now we explain what is happening in the Fig. 1.

For example in Fig.~\ref{fig1}, at  $\theta =0^\circ $, the Hamiltonian comprises the hyperfine and z-component of the 
Zeeman interaction; therefore, in accordance with (i) and (iii), the  spin  evolution corresponds to the coherent 
mixing between $\ket{s} \leftrightarrow \ket{t_0}$ and the singlet yield saturates at 0.50 for larger hyperfine
interaction which corresponds to strong mixing within the recombination timescale.
On the other extreme,  at  $\theta = 90^\circ$, the Hamiltonian comprises the hyperfine interaction and 
x-component of the Zeeman interaction; based on (i) and (ii), spin transition pathways now exist between all 
the four spin states by the collective effect of the two interactions and the yields corresponding to 
all spin states saturate to 0.25 for larger hyperfine interaction strength.
Thus, the compass action in the RP model is a consequence of the collective spin dynamical behavior due to the hyperfine 
and Zeeman interactions. 
\begin{myfigure}[t]
\includegraphics[width=6.5cm,height=6.5cm,keepaspectratio]{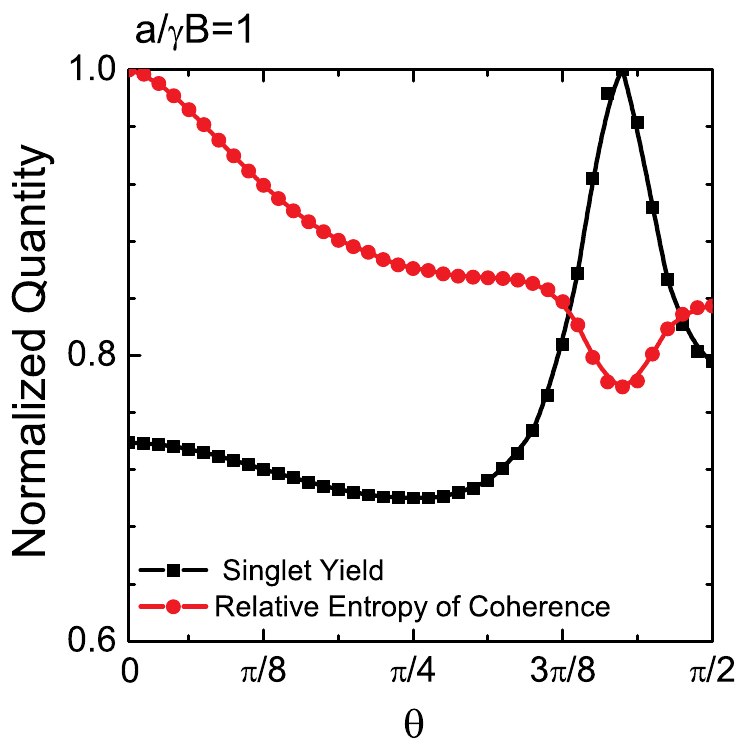}
\caption{The normalized singlet yield (black) and the normalized relative entropy of coherence (red) as a function of 
magnetic field orientation $\theta$ at $a/\gamma B_0=1 $. \iffalse along with relative entropy of coherence as a function of magnetic field inclination. \fi 
The \textit{relative entropy of coherence} serves as a measure of coherence~\citep{baumgratz2014quantifying}. 
The dip in coherence corresponding to the peak in singlet yield indicates that the peaks in 
singlet yield in Fig.~\ref{fig1} are due to nuclear decoherence. This fact is further affirmed by the singlet yield curve obtained when the nuclear and the radical 
pair spin states are disentangled, cf. Appendix~\ref{Appen_NuclearDecoherence}.}
\label{fig3}
\end{myfigure}
\section{Nuclear decoherence}
\label{NuclearDecoherence}
From Fig.~\ref{fig1}, we also observe that there is a conspicuous peak appearing in the singlet yield whose 
position depends on the hyperfine coupling strength; there are corresponding dips in $\ket{T_+}$ and $\ket{T_-}$ yields 
suggesting a mechanism that blocks the spin transition pathways from  $\ket{s}$ to $\ket{t_+}$ and $\ket{t_-}$. 
We find that the singlet yield peak vanishes when we disentangle
the nuclear and electronic spin dynamics while retaining the effect of the hyperfine interaction,
indicating that the peak is actually due to nuclear decoherence. 
Using a quantum information theoretic quantifier of coherence -- the relative entropy of 
coherence~\citep{baumgratz2014quantifying} -- we seek to establish a connection between the singlet 
yield peak and nuclear decoherence. The coherence is characterized in the computational 
basis of the three  spin system.  If $\rho$ is the density matrix  of the system, the relative entropy of coherence 
is given by~\citep{baumgratz2014quantifying}:
\begin{eqnarray}
\label{eq3}
C(\rho) \equiv S(\rho_{diag})-S(\rho)
\end{eqnarray}
where $S(\rho)$ is von Neumann entropy corresponding to $\rho$ and $\rho_{diag}$ is obtained by taking only 
diagonal elements of $\rho$. The  normalized  singlet  yield and  normalized  relative entropy of coherence are 
plotted as a function of magnetic field orientation in Fig.~\ref{fig3}. 
We see a dip in the  relative entropy of coherence corresponding to the peak in singlet 
yield which corroborates our hypothesis that the peak in singlet yield is due to nuclear decoherence. 

Defining the sensitivity as: $D_S=\Phi_S^{max} - \Phi_S^{min}$, where $\Phi_S$ is the singlet yield and the max/min
are with respect to the magnetic field orientation, the  aforementioned analysis
of singlet yield for various  hyperfine  interaction strengths reproduces the sensitivity behavior 
shown by J. Cai et al.~\citep{cai2012quantum}.
The ripples in their sensitivity vs.  $a/ \gamma B$ plot (Fig. 1 in that paper) around $a/ \gamma B = 1$ can now be 
understood as a consequence of peaks in the singlet yield and thus a direct manifestation of the nuclear decoherence.
Other results reported in that work can also be interpreted along similar lines -- that is, by considering the spin state
transitions induced by each of the Hamiltonian terms.
\begin{myfigure}[t]
\includegraphics[]{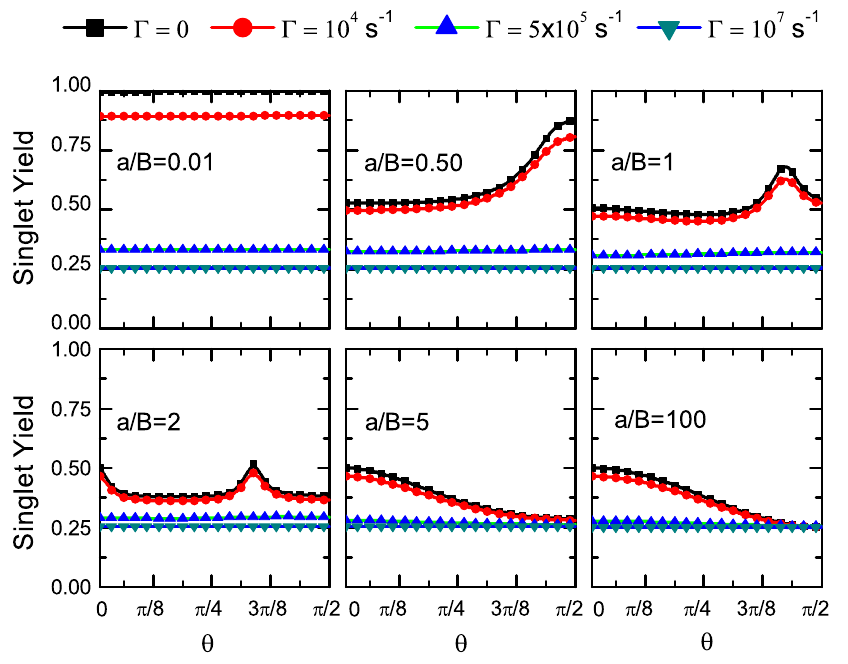}
\caption{Singlet yield as a function of geomagnetic orientation in presence of environmental noise for the noise rates of 
$\Gamma = $ $0$ $s^{-1}$, $10^4$ $s^{-1}$, $5 \times 10^5$ $s^{-1}$, $10^7$ $s^{-1}$. The figure shows that if 
noise rate is equal to or greater than $k$, the orientation dependence of the singlet yield is destroyed.}
\label{fig4}
\end{myfigure}
\section{Environmental decoherence}
\label{EnvironDecoherence}
We adopt the environmental noise model from Gauger et al.~\citep{gauger2011sustained}.
Mathematically, we consider the following six noise operators: $L_1 = I_2 \otimes \sigma_x \otimes I_2 $, 
$L_2 = I_2 \otimes \sigma_y \otimes I_2 $, $L_3 = I_2 \otimes \sigma_z \otimes I_2 $, 
$L_4 = I_2 \otimes I_2 \otimes \sigma_x $, $L_5 = I_2 \otimes I_2 \otimes \sigma_y $, 
$L_6 = I_2 \otimes I_2 \otimes \sigma_z $. The master equation now gets modified as: 
$\dot \rho = - \frac{i}{\hbar}[H, \rho] + k \sum\limits_{i=1}^8 P_i \rho P_i^\dagger - \frac{1}{2}(P_i^\dagger P_i \rho + \rho P_i^\dagger P_i) + 
\Gamma \sum\limits_{i=1}^6 L_i \rho L_i^\dagger - \frac{1}{2}(L_i^\dagger L_i \rho + \rho L_i^\dagger L_i)$, where
$\Gamma$ is the noise rate~\citep{gauger2011sustained}.
Fig.~\ref{fig4} shows the singlet yield for four different noise rates, calculated for six different hyperfine coupling
strengths. Expectedly, the effect of the environmental noise is most substantial when the noise rate is equal to or 
greater than the recombination rate \textit{k}; in these cases, we find that the compass sensitivity goes to zero.
This  indicates that environmental noise opens the spin transition pathways between the singlet state 
and all triplet states and thereby homogenizing the yields corresponding to all the spin states. 
Therefore, the singlet yield saturates to 0.25, as shown for the case in which the noise rate ($\Gamma$) is 
greater than $k$.
\section{Conclusion}
\label{Conclusion}
In this study,  we elucidate the essential spin dynamics of the radical pair model for avian magnetoreception and 
inspect the distinct effects of nuclear and environmental decoherence therein. We find that the Zeeman and hyperfine 
interactions play collaborative roles to enable the compass action. In particular, we have identified the dual role played
by the nucleus: while the anisotropic nuclear interaction is necessary to induce spin transitions that make
the RP dynamics magneto-sensitive, decoherence due to the nucleus introduces additional structure in the dynamics
but does not affect the compass action.
On the other hand, decoherence due to the environment tends to make the singlet yield insensitive to the geomagnetic 
field orientation by opening transition pathways between all the spin states; as expected, this destroys the compass action.
We think that our approach of understanding the sensitivity of the avian compass -- in terms of spin-state transitions 
induced by each of the Hamiltonian terms, and its limitation -- in terms of nuclear and environmental decoherence, 
may suggest guidelines for its solid state emulation. The diamond nitrogen vacancy center spin system seems 
to be a potential solid state candidate for this purpose owing to its room temperature long coherence time and 
two spin relaxation pathways~\citep{doherty2013nitrogen,maze2008nanoscale,balasubramanian2009ultralong}. 
However, a lot of work remains to be done before this, or any other solid state system, can emulate the avian compass
and thereby open the path towards geomagnetic field assisted navigation systems.
\acknowledgments{We are highly grateful to E. M. Gauger, J. Cai, and M. B. Plenio for insightful communications.}
\appendix
%\label{Appendices}
\begin{table*}[t]
\setlength{\tabcolsep}{12pt}
\begin{tabular}{|l|l|l|l|l|l|l|}
\hline
\multicolumn{3}{|c|}{Hyperfine Interactions}&\multicolumn{3}{c|}{Zeeman Interactions}\\
\cline{1-6}
 x-hyperfine  & y-hyperfine  &z-hyperfine  &x-Zeeman  &y-Zeeman  &z-Zeeman\\
\hline
$S \leftrightarrow T_{+}, T_{-}$ & $S \leftrightarrow T_{+}, T_{-}$ & $S \leftrightarrow S, T_{0}$ & $S \leftrightarrow S$ & $S \leftrightarrow S$ & $S \leftrightarrow S$\\
$T_{0} \leftrightarrow T_{+}, T_{-}$ & $T_{0} \leftrightarrow T_{+}, T_{-}$ & $T_{0} \leftrightarrow S, T_{0}$ & $T_{0} \leftrightarrow T_{+}, T_{-}$ & $T_{0} \leftrightarrow T_{+}, T_{-}$ & $T_{0} \leftrightarrow T_{0}$\\
$T_{+} \leftrightarrow T_{0}, S$ & $T_{+} \leftrightarrow T_{0}, S$ & $T_{+} \leftrightarrow T_{+}$ & $T_{+} \leftrightarrow T_{0}$ & $T_{+} \leftrightarrow T_{0}$ & $T_{+} \leftrightarrow T_{+}$\\
$T_{-} \leftrightarrow T_{0}, S$ & $T_{-} \leftrightarrow T_{0}, S$ & $T_{-} \leftrightarrow T_{-}$ & $T_{-} \leftrightarrow T_{0}$ & $T_{-} \leftrightarrow T_{0}$ & $T_{-} \leftrightarrow T_{-}$ \\
\hline
\end{tabular}
\caption{Spin transitions induced by hyperfine and Zeeman interactions.}
\label{SpinTransitions}
\end{table*}
\section{Spin Transitions due to Hyperfine and Zeeman Interactions}
\label{Appen_SpinTransitionsRP}
The radical pair Hamiltonian is given in Eq.~\ref{eq1}.
Under this Hamiltonian, the spin evolution of joint system (radical pair + nucleus) is given by Eq.~\ref{eq2}.
Here, $\rho = \rho_{nuc} \otimes \rho_{RP}$ is the state of the joint system. At t=0, the state of the joint system is
given as: $\rho(0) = \frac{1}{2}I \otimes (\ket{s} \otimes \bra{s})$. 

In order to understand the spin transitions induced by each Hamiltonian interaction, we calculate the density matrix evolution under these Hamiltonian terms.
If the Hamiltonian is time independent (which is the case here), the evolution of the density matrix is given as: 
\begin{eqnarray}
\label{Supp_eq1}
\rho(t) = e^{\frac{-iHt}{\hbar}}\rho(0)e^{\frac{iHt}{\hbar}}
\end{eqnarray}
Under hyperfine part of the Hamiltonian, $H_{hyp}=\hat{I} \cdot \mathbf{A} \cdot\hat{S_2}$, the state of the system evolves as:
\begin{eqnarray}
\label{Supp_eq2}
\rho(t) = e^{\frac{-iH_{hyp}t}{\hbar}}\rho(0)e^{\frac{iH_{hyp}t}{\hbar}}
\end{eqnarray}
By tracing out the nuclear part, we get the radical pair state of the Hamiltonian i.e., $\rho_{RP}(t) = tr_{nuc}(\rho(t))$.
Similarly, we can calculate how the radical pair state changes under Zeeman Hamiltonian, $H_{Zeeman}=\gamma \mathbf{B} \cdot (\hat{S_1} + \hat{S_2})$.
The results of these calculations are summarized in Table.~\ref{SpinTransitions}. These results are also in agreement with the results 
of B. M. Xu et al.~\citep{xu2014effect}.
\begin{myfigure}[htbp]
\includegraphics[width=8cm,height=8cm,keepaspectratio]{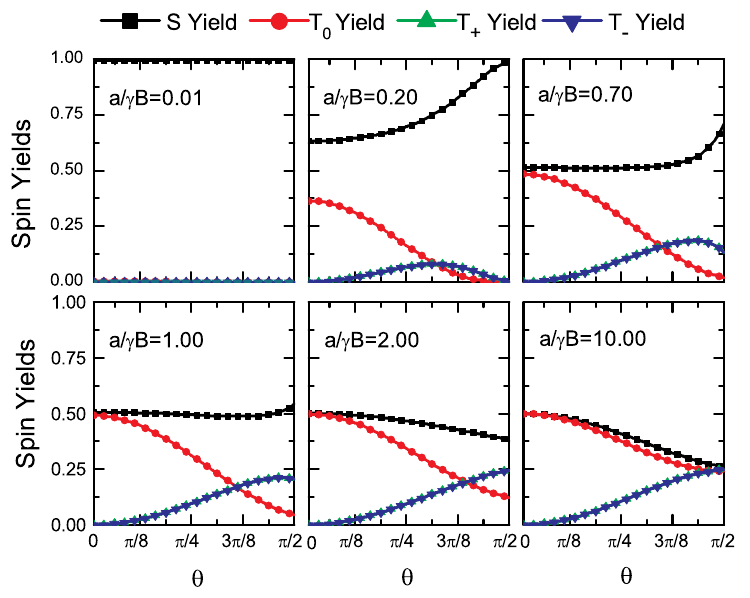}
\caption{Singlet ($S$) and triplet ($T_0$, $T_+$, $T_-$) yields vs. geomagnetic field orientation for various hyperfine coupling strengths. 
Now the spin state of the nucleus and radical pair are disentangled. Unlike Fig.~\ref{fig1}, the peaks in singlet yield do not appear 
here. This confirms that the peaks in singlet yield (Fig.~\ref{fig1}) are direct manifestation of nuclear decoherence.} 
\label{SFig1}
\end{myfigure}
\begin{myfigure}[htbp]
\includegraphics[width=7cm,height=7cm,keepaspectratio]{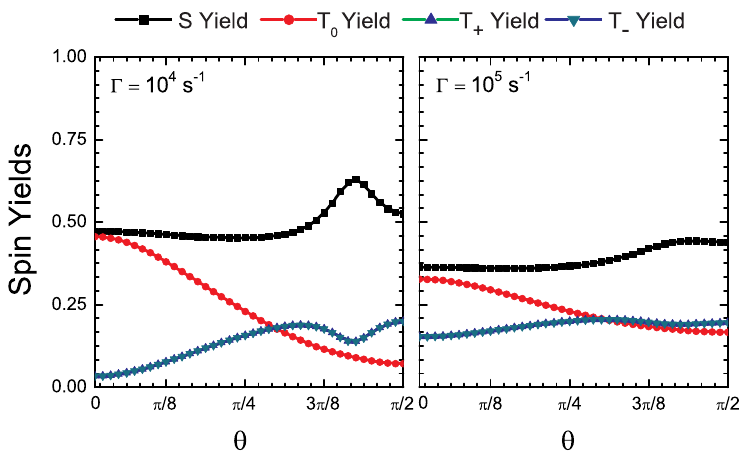} 
\caption{Singlet ($S$) and Triplet ($T_0$, $T_+$, $T_-$) yields vs geomagnetic field orientation in 
presence of environmental noise with noise rate ($\Gamma$) = $10^4 s^{-1}$ and $10^5 s^{-1}$. 
The figures shows that spin yields tend to become uniform as the environmental noise rate is increased. 
Thus environmental noise opens the spin transition 
pathways between all four spin states of the radical pair.} 
\label{SFig2}
\end{myfigure}

\section{Nuclear Decoherence}
\label{Appen_NuclearDecoherence}
The singlet yield shown in Fig.~\ref{fig1} shows a peak in the curve. The peaks are hypothesized to be due to nuclear decoherence
which is also verified by quantifying the nuclear decoherence by relative entropy~\citep{baumgratz2014quantifying}, 
shown in Fig.~\ref{fig3}. Here, we present yet another evidence that the singlet yield peaks are the consequence of nuclear decoherence. 
We calculate the singlet yield by disentangling the radical pair state from the nuclear state
i.e. instead of taking hyperfine interaction as $H_{hf} = \hat{I} \cdot \mathbf{A} \cdot\hat{S_2} = 
\sum\limits_{i=x,y,z} \sigma_i \otimes I \otimes \sigma_i$, 
we consider the ~\textit{modified hyperfine interaction}: $H_{mhf} = \sum\limits_{i=x,y,z} I \otimes I \otimes \sigma_i$; this eliminates the evolution of the 
nuclear state and effectively treats the nuclear spin as a source of static magnetic field acting on one of the RP spins. Fig.~\ref{SFig1} shows the 
singlet yield for a number of modified hyperfine interaction strengths. 
Comparing with Fig.~\ref{fig1}, we can clearly see that now the peaks in singlet yield have disappeared, indicating that 
these are due to nuclear decoherence.
\section{Effect of Environmental Decoherence on All Spin Yields}
\label{Appen_EnvironDecoherence}
The effect of environmental noise on singlet yield has been shown in Fig.~\ref{fig4}. Fig.~\ref{SFig2} displays the 
effect of environmental noise on all spin yields ($S,T_0,T_+,T_-$) for environmental noise rates of $10^4 s^{-1}$ and $10^{5} s^{-1}$. Here we observe 
that as the noise rate is increased, the yield corresponding to all spins tends to converge to 0.25. Thus, this plot establishes the fact that environmental noise 
opens up the spin transition pathways between the singlet state and all triplet states and thereby homogenizing the yields corresponding to all the spin 
states. Thus environmental noise washes away the sensitivity of the avian compass.

\bibliography{State_Transitions_and_Decoherence_in_the_Avian_Compass}

\providecommand{\noopsort}[1]{}\providecommand{\singleletter}[1]{#1}%
\begin{thebibliography}{36}%
\makeatletter
\providecommand \@ifxundefined [1]{%
 \@ifx{#1\undefined}
}%
\providecommand \@ifnum [1]{%
 \ifnum #1\expandafter \@firstoftwo
 \else \expandafter \@secondoftwo
 \fi
}%
\providecommand \@ifx [1]{%
 \ifx #1\expandafter \@firstoftwo
 \else \expandafter \@secondoftwo
 \fi
}%
\providecommand \natexlab [1]{#1}%
\providecommand \enquote  [1]{``#1''}%
\providecommand \bibnamefont  [1]{#1}%
\providecommand \bibfnamefont [1]{#1}%
\providecommand \citenamefont [1]{#1}%
\providecommand \href@noop [0]{\@secondoftwo}%
\providecommand \href [0]{\begingroup \@sanitize@url \@href}%
\providecommand \@href[1]{\@@startlink{#1}\@@href}%
\providecommand \@@href[1]{\endgroup#1\@@endlink}%
\providecommand \@sanitize@url [0]{\catcode `\\12\catcode `\$12\catcode
  `\&12\catcode `\#12\catcode `\^12\catcode `\_12\catcode `\%12\relax}%
\providecommand \@@startlink[1]{}%
\providecommand \@@endlink[0]{}%
\providecommand \url  [0]{\begingroup\@sanitize@url \@url }%
\providecommand \@url [1]{\endgroup\@href {#1}{\urlprefix }}%
\providecommand \urlprefix  [0]{URL }%
\providecommand \Eprint [0]{\href }%
\providecommand \doibase [0]{http://dx.doi.org/}%
\providecommand \selectlanguage [0]{\@gobble}%
\providecommand \bibinfo  [0]{\@secondoftwo}%
\providecommand \bibfield  [0]{\@secondoftwo}%
\providecommand \translation [1]{[#1]}%
\providecommand \BibitemOpen [0]{}%
\providecommand \bibitemStop [0]{}%
\providecommand \bibitemNoStop [0]{.\EOS\space}%
\providecommand \EOS [0]{\spacefactor3000\relax}%
\providecommand \BibitemShut  [1]{\csname bibitem#1\endcsname}%
\let\auto@bib@innerbib\@empty
%</preamble>
\bibitem [{\citenamefont {Engel}\ \emph {et~al.}(2007)\citenamefont {Engel},
  \citenamefont {Calhoun}, \citenamefont {Read}, \citenamefont {Ahn},
  \citenamefont {Man{\v{c}}al}, \citenamefont {Cheng}, \citenamefont
  {Blankenship},\ and\ \citenamefont {Fleming}}]{engel2007evidence}%
  \BibitemOpen
  \bibfield  {author} {\bibinfo {author} {\bibfnamefont {G.~S.}\ \bibnamefont
  {Engel}}, \bibinfo {author} {\bibfnamefont {T.~R.}\ \bibnamefont {Calhoun}},
  \bibinfo {author} {\bibfnamefont {E.~L.}\ \bibnamefont {Read}}, \bibinfo
  {author} {\bibfnamefont {T.-K.}\ \bibnamefont {Ahn}}, \bibinfo {author}
  {\bibfnamefont {T.}~\bibnamefont {Man{\v{c}}al}}, \bibinfo {author}
  {\bibfnamefont {Y.-C.}\ \bibnamefont {Cheng}}, \bibinfo {author}
  {\bibfnamefont {R.~E.}\ \bibnamefont {Blankenship}}, \ and\ \bibinfo {author}
  {\bibfnamefont {G.~R.}\ \bibnamefont {Fleming}},\ }\href
  {http://www.nature.com/nature/journal/v446/n7137/full/nature05678.html}
  {\bibfield  {journal} {\bibinfo  {journal} {Nature}\ }\textbf {\bibinfo
  {volume} {446}},\ \bibinfo {pages} {782} (\bibinfo {year}
  {2007})}\BibitemShut {NoStop}%
\bibitem [{\citenamefont {Plenio}\ and\ \citenamefont
  {Huelga}(2008)}]{plenio2008dephasing}%
  \BibitemOpen
  \bibfield  {author} {\bibinfo {author} {\bibfnamefont {M.~B.}\ \bibnamefont
  {Plenio}}\ and\ \bibinfo {author} {\bibfnamefont {S.~F.}\ \bibnamefont
  {Huelga}},\ }\href {http://iopscience.iop.org/1367-2630/10/11/113019}
  {\bibfield  {journal} {\bibinfo  {journal} {New Journal of Physics}\ }\textbf
  {\bibinfo {volume} {10}},\ \bibinfo {pages} {113019} (\bibinfo {year}
  {2008})}\BibitemShut {NoStop}%
\bibitem [{\citenamefont {Lambert}\ \emph {et~al.}(2013)\citenamefont
  {Lambert}, \citenamefont {Chen}, \citenamefont {Cheng}, \citenamefont {Li},
  \citenamefont {Chen},\ and\ \citenamefont {Nori}}]{lambert2013quantum}%
  \BibitemOpen
  \bibfield  {author} {\bibinfo {author} {\bibfnamefont {N.}~\bibnamefont
  {Lambert}}, \bibinfo {author} {\bibfnamefont {Y.-N.}\ \bibnamefont {Chen}},
  \bibinfo {author} {\bibfnamefont {Y.-C.}\ \bibnamefont {Cheng}}, \bibinfo
  {author} {\bibfnamefont {C.-M.}\ \bibnamefont {Li}}, \bibinfo {author}
  {\bibfnamefont {G.-Y.}\ \bibnamefont {Chen}}, \ and\ \bibinfo {author}
  {\bibfnamefont {F.}~\bibnamefont {Nori}},\ }\href
  {http://www.nature.com/nphys/journal/v9/n1/abs/nphys2474.html} {\bibfield
  {journal} {\bibinfo  {journal} {Nature Physics}\ }\textbf {\bibinfo {volume}
  {9}},\ \bibinfo {pages} {10} (\bibinfo {year} {2013})}\BibitemShut {NoStop}%
\bibitem [{\citenamefont {Mohseni}\ \emph {et~al.}(2008)\citenamefont
  {Mohseni}, \citenamefont {Rebentrost}, \citenamefont {Lloyd},\ and\
  \citenamefont {Aspuru-Guzik}}]{mohseni2008environment}%
  \BibitemOpen
  \bibfield  {author} {\bibinfo {author} {\bibfnamefont {M.}~\bibnamefont
  {Mohseni}}, \bibinfo {author} {\bibfnamefont {P.}~\bibnamefont {Rebentrost}},
  \bibinfo {author} {\bibfnamefont {S.}~\bibnamefont {Lloyd}}, \ and\ \bibinfo
  {author} {\bibfnamefont {A.}~\bibnamefont {Aspuru-Guzik}},\ }\href
  {http://scitation.aip.org/content/aip/journal/jcp/129/17/10.1063/1.3002335}
  {\bibfield  {journal} {\bibinfo  {journal} {The Journal of chemical physics}\
  }\textbf {\bibinfo {volume} {129}},\ \bibinfo {pages} {174106} (\bibinfo
  {year} {2008})}\BibitemShut {NoStop}%
\bibitem [{\citenamefont {Scholes}\ \emph {et~al.}(2011)\citenamefont
  {Scholes}, \citenamefont {Fleming}, \citenamefont {Olaya-Castro},\ and\
  \citenamefont {van Grondelle}}]{scholes2011lessons}%
  \BibitemOpen
  \bibfield  {author} {\bibinfo {author} {\bibfnamefont {G.~D.}\ \bibnamefont
  {Scholes}}, \bibinfo {author} {\bibfnamefont {G.~R.}\ \bibnamefont
  {Fleming}}, \bibinfo {author} {\bibfnamefont {A.}~\bibnamefont
  {Olaya-Castro}}, \ and\ \bibinfo {author} {\bibfnamefont {R.}~\bibnamefont
  {van Grondelle}},\ }\href
  {http://www.nature.com/nchem/journal/v3/n10/full/nchem.1145.html} {\bibfield
  {journal} {\bibinfo  {journal} {Nature chemistry}\ }\textbf {\bibinfo
  {volume} {3}},\ \bibinfo {pages} {763} (\bibinfo {year} {2011})}\BibitemShut
  {NoStop}%
\bibitem [{\citenamefont {Turin}(1996)}]{turin1996spectroscopic}%
  \BibitemOpen
  \bibfield  {author} {\bibinfo {author} {\bibfnamefont {L.}~\bibnamefont
  {Turin}},\ }\href
  {http://chemse.oxfordjournals.org/content/21/6/773.abstract} {\bibfield
  {journal} {\bibinfo  {journal} {Chemical Senses}\ }\textbf {\bibinfo {volume}
  {21}},\ \bibinfo {pages} {773} (\bibinfo {year} {1996})}\BibitemShut
  {NoStop}%
\bibitem [{\citenamefont {Brookes}\ \emph {et~al.}(2007)\citenamefont
  {Brookes}, \citenamefont {Hartoutsiou}, \citenamefont {Horsfield},\ and\
  \citenamefont {Stoneham}}]{brookes2007could}%
  \BibitemOpen
  \bibfield  {author} {\bibinfo {author} {\bibfnamefont {J.~C.}\ \bibnamefont
  {Brookes}}, \bibinfo {author} {\bibfnamefont {F.}~\bibnamefont
  {Hartoutsiou}}, \bibinfo {author} {\bibfnamefont {A.}~\bibnamefont
  {Horsfield}}, \ and\ \bibinfo {author} {\bibfnamefont {A.}~\bibnamefont
  {Stoneham}},\ }\href
  {http://journals.aps.org/prl/abstract/10.1103/PhysRevLett.98.038101}
  {\bibfield  {journal} {\bibinfo  {journal} {Physical review letters}\
  }\textbf {\bibinfo {volume} {98}},\ \bibinfo {pages} {038101} (\bibinfo
  {year} {2007})}\BibitemShut {NoStop}%
\bibitem [{\citenamefont {Nagel}\ and\ \citenamefont
  {Klinman}(2006)}]{nagel2006tunneling}%
  \BibitemOpen
  \bibfield  {author} {\bibinfo {author} {\bibfnamefont {Z.~D.}\ \bibnamefont
  {Nagel}}\ and\ \bibinfo {author} {\bibfnamefont {J.~P.}\ \bibnamefont
  {Klinman}},\ }\href {http://pubs.acs.org/doi/pdf/10.1021/cr050301x}
  {\bibfield  {journal} {\bibinfo  {journal} {Chemical reviews}\ }\textbf
  {\bibinfo {volume} {106}},\ \bibinfo {pages} {3095} (\bibinfo {year}
  {2006})}\BibitemShut {NoStop}%
\bibitem [{\citenamefont {Wiltschko}\ and\ \citenamefont
  {Wiltschko}(1972)}]{wiltschko1972magnetic}%
  \BibitemOpen
  \bibfield  {author} {\bibinfo {author} {\bibfnamefont {W.}~\bibnamefont
  {Wiltschko}}\ and\ \bibinfo {author} {\bibfnamefont {R.}~\bibnamefont
  {Wiltschko}},\ }\href {http://www.sciencemag.org/content/176/4030/62.short}
  {\bibfield  {journal} {\bibinfo  {journal} {Science}\ }\textbf {\bibinfo
  {volume} {176}},\ \bibinfo {pages} {62} (\bibinfo {year} {1972})}\BibitemShut
  {NoStop}%
\bibitem [{\citenamefont {Schulten}\ \emph {et~al.}(1978)\citenamefont
  {Schulten}, \citenamefont {Swenberg},\ and\ \citenamefont
  {Weller}}]{schulten1978biomagnetic}%
  \BibitemOpen
  \bibfield  {author} {\bibinfo {author} {\bibfnamefont {K.}~\bibnamefont
  {Schulten}}, \bibinfo {author} {\bibfnamefont {C.~E.}\ \bibnamefont
  {Swenberg}}, \ and\ \bibinfo {author} {\bibfnamefont {A.}~\bibnamefont
  {Weller}},\ }\href
  {http://www.degruyter.com/view/j/zpch.1978.111.issue-1/zpch.1978.111.1.001/zpch.1978.111.1.001.xml}
  {\bibfield  {journal} {\bibinfo  {journal} {Z. Phys. Chem}\ }\textbf
  {\bibinfo {volume} {111}},\ \bibinfo {pages} {1} (\bibinfo {year}
  {1978})}\BibitemShut {NoStop}%
\bibitem [{\citenamefont {Wiltschko}\ \emph {et~al.}(2010)\citenamefont
  {Wiltschko}, \citenamefont {Stapput}, \citenamefont {Thalau},\ and\
  \citenamefont {Wiltschko}}]{wiltschko2010directional}%
  \BibitemOpen
  \bibfield  {author} {\bibinfo {author} {\bibfnamefont {R.}~\bibnamefont
  {Wiltschko}}, \bibinfo {author} {\bibfnamefont {K.}~\bibnamefont {Stapput}},
  \bibinfo {author} {\bibfnamefont {P.}~\bibnamefont {Thalau}}, \ and\ \bibinfo
  {author} {\bibfnamefont {W.}~\bibnamefont {Wiltschko}},\ }\href
  {http://www.ncbi.nlm.nih.gov/pubmed/19864263} {\bibfield  {journal} {\bibinfo
   {journal} {Journal of The Royal Society Interface}\ }\textbf {\bibinfo
  {volume} {7}},\ \bibinfo {pages} {S163} (\bibinfo {year} {2010})}\BibitemShut
  {NoStop}%
\bibitem [{\citenamefont {Steiner}\ and\ \citenamefont
  {Ulrich}(1989)}]{steiner1989magnetic}%
  \BibitemOpen
  \bibfield  {author} {\bibinfo {author} {\bibfnamefont {U.~E.}\ \bibnamefont
  {Steiner}}\ and\ \bibinfo {author} {\bibfnamefont {T.}~\bibnamefont
  {Ulrich}},\ }\href {http://pubs.acs.org/doi/abs/10.1021/cr00091a003}
  {\bibfield  {journal} {\bibinfo  {journal} {Chemical Reviews}\ }\textbf
  {\bibinfo {volume} {89}},\ \bibinfo {pages} {51} (\bibinfo {year}
  {1989})}\BibitemShut {NoStop}%
\bibitem [{\citenamefont {Ritz}\ \emph {et~al.}(2000)\citenamefont {Ritz},
  \citenamefont {Adem},\ and\ \citenamefont {Schulten}}]{ritz2000model}%
  \BibitemOpen
  \bibfield  {author} {\bibinfo {author} {\bibfnamefont {T.}~\bibnamefont
  {Ritz}}, \bibinfo {author} {\bibfnamefont {S.}~\bibnamefont {Adem}}, \ and\
  \bibinfo {author} {\bibfnamefont {K.}~\bibnamefont {Schulten}},\ }\href
  {http://www.ncbi.nlm.nih.gov/pmc/articles/PMC1300674/} {\bibfield  {journal}
  {\bibinfo  {journal} {Biophysical journal}\ }\textbf {\bibinfo {volume}
  {78}},\ \bibinfo {pages} {707} (\bibinfo {year} {2000})}\BibitemShut
  {NoStop}%
\bibitem [{\citenamefont {Solov'yov}\ and\ \citenamefont
  {Schulten}(2009)}]{solov2009magnetoreception}%
  \BibitemOpen
  \bibfield  {author} {\bibinfo {author} {\bibfnamefont {I.~A.}\ \bibnamefont
  {Solov'yov}}\ and\ \bibinfo {author} {\bibfnamefont {K.}~\bibnamefont
  {Schulten}},\ }\href {http://www.ncbi.nlm.nih.gov/pubmed/19527640} {\bibfield
   {journal} {\bibinfo  {journal} {Biophysical journal}\ }\textbf {\bibinfo
  {volume} {96}},\ \bibinfo {pages} {4804} (\bibinfo {year}
  {2009})}\BibitemShut {NoStop}%
\bibitem [{\citenamefont {Solov’yov}\ and\ \citenamefont
  {Schulten}(2012)}]{solov2012reaction}%
  \BibitemOpen
  \bibfield  {author} {\bibinfo {author} {\bibfnamefont {I.~A.}\ \bibnamefont
  {Solov’yov}}\ and\ \bibinfo {author} {\bibfnamefont {K.}~\bibnamefont
  {Schulten}},\ }\href {http://pubs.acs.org/doi/abs/10.1021/jp209508y}
  {\bibfield  {journal} {\bibinfo  {journal} {The Journal of Physical Chemistry
  B}\ }\textbf {\bibinfo {volume} {116}},\ \bibinfo {pages} {1089} (\bibinfo
  {year} {2012})}\BibitemShut {NoStop}%
\bibitem [{\citenamefont {Timmel}\ and\ \citenamefont
  {Henbest}(2004)}]{timmel2004study}%
  \BibitemOpen
  \bibfield  {author} {\bibinfo {author} {\bibfnamefont {C.~R.}\ \bibnamefont
  {Timmel}}\ and\ \bibinfo {author} {\bibfnamefont {K.~B.}\ \bibnamefont
  {Henbest}},\ }\href
  {http://rsta.royalsocietypublishing.org/content/362/1825/2573} {\bibfield
  {journal} {\bibinfo  {journal} {Philosophical Transactions of the Royal
  Society of London. Series A: Mathematical, Physical and Engineering
  Sciences}\ }\textbf {\bibinfo {volume} {362}},\ \bibinfo {pages} {2573}
  (\bibinfo {year} {2004})}\BibitemShut {NoStop}%
\bibitem [{\citenamefont {Miura}\ \emph {et~al.}(2006)\citenamefont {Miura},
  \citenamefont {Maeda},\ and\ \citenamefont {Arai}}]{miura2006spin}%
  \BibitemOpen
  \bibfield  {author} {\bibinfo {author} {\bibfnamefont {T.}~\bibnamefont
  {Miura}}, \bibinfo {author} {\bibfnamefont {K.}~\bibnamefont {Maeda}}, \ and\
  \bibinfo {author} {\bibfnamefont {T.}~\bibnamefont {Arai}},\ }\href
  {http://pubs.acs.org/doi/abs/10.1021/jp056488d} {\bibfield  {journal}
  {\bibinfo  {journal} {The Journal of Physical Chemistry A}\ }\textbf
  {\bibinfo {volume} {110}},\ \bibinfo {pages} {4151} (\bibinfo {year}
  {2006})}\BibitemShut {NoStop}%
\bibitem [{\citenamefont {Rodgers}(2009)}]{rodgers2009magnetic}%
  \BibitemOpen
  \bibfield  {author} {\bibinfo {author} {\bibfnamefont {C.~T.}\ \bibnamefont
  {Rodgers}},\ }\href
  {http://www.degruyter.com/view/j/pac.2009.81.issue-1/pac-con-08-10-18/pac-con-08-10-18.xml}
  {\bibfield  {journal} {\bibinfo  {journal} {Pure \& Applied Chemistry}\
  }\textbf {\bibinfo {volume} {81}} (\bibinfo {year} {2009})}\BibitemShut
  {NoStop}%
\bibitem [{\citenamefont {Lau}(2014)}]{lau2014spin}%
  \BibitemOpen
  \bibfield  {author} {\bibinfo {author} {\bibfnamefont {J.}~\bibnamefont
  {Lau}},\ }\emph {\bibinfo {title} {Spin-selective chemical reactions in
  radical pair magnetoreception}},\ \href
  {http://ethos.bl.uk/OrderDetails.do?uin=uk.bl.ethos.604508} {Ph.D. thesis},\
  \bibinfo  {school} {University of Oxford} (\bibinfo {year}
  {2014})\BibitemShut {NoStop}%
\bibitem [{\citenamefont {Wiltschko}\ \emph {et~al.}(1993)\citenamefont
  {Wiltschko}, \citenamefont {Munro}, \citenamefont {Ford},\ and\ \citenamefont
  {Wiltschko}}]{wiltschko1993red}%
  \BibitemOpen
  \bibfield  {author} {\bibinfo {author} {\bibfnamefont {W.}~\bibnamefont
  {Wiltschko}}, \bibinfo {author} {\bibfnamefont {U.}~\bibnamefont {Munro}},
  \bibinfo {author} {\bibfnamefont {H.}~\bibnamefont {Ford}}, \ and\ \bibinfo
  {author} {\bibfnamefont {R.}~\bibnamefont {Wiltschko}},\ }\href
  {http://www.nature.com/nature/journal/v364/n6437/abs/364525a0.html} {\
  (\bibinfo {year} {1993})}\BibitemShut {NoStop}%
\bibitem [{\citenamefont {Ritz}\ \emph {et~al.}(2009)\citenamefont {Ritz},
  \citenamefont {Wiltschko}, \citenamefont {Hore}, \citenamefont {Rodgers},
  \citenamefont {Stapput}, \citenamefont {Thalau}, \citenamefont {Timmel},\
  and\ \citenamefont {Wiltschko}}]{ritz2009magnetic}%
  \BibitemOpen
  \bibfield  {author} {\bibinfo {author} {\bibfnamefont {T.}~\bibnamefont
  {Ritz}}, \bibinfo {author} {\bibfnamefont {R.}~\bibnamefont {Wiltschko}},
  \bibinfo {author} {\bibfnamefont {P.}~\bibnamefont {Hore}}, \bibinfo {author}
  {\bibfnamefont {C.~T.}\ \bibnamefont {Rodgers}}, \bibinfo {author}
  {\bibfnamefont {K.}~\bibnamefont {Stapput}}, \bibinfo {author} {\bibfnamefont
  {P.}~\bibnamefont {Thalau}}, \bibinfo {author} {\bibfnamefont {C.~R.}\
  \bibnamefont {Timmel}}, \ and\ \bibinfo {author} {\bibfnamefont
  {W.}~\bibnamefont {Wiltschko}},\ }\href
  {http://www.ncbi.nlm.nih.gov/pmc/articles/PMC2718301/} {\bibfield  {journal}
  {\bibinfo  {journal} {Biophysical journal}\ }\textbf {\bibinfo {volume}
  {96}},\ \bibinfo {pages} {3451} (\bibinfo {year} {2009})}\BibitemShut
  {NoStop}%
\bibitem [{\citenamefont {Tiersch}\ and\ \citenamefont
  {Briegel}(2012)}]{tiersch2012decoherence}%
  \BibitemOpen
  \bibfield  {author} {\bibinfo {author} {\bibfnamefont {M.}~\bibnamefont
  {Tiersch}}\ and\ \bibinfo {author} {\bibfnamefont {H.~J.}\ \bibnamefont
  {Briegel}},\ }\href
  {http://rsta.royalsocietypublishing.org/content/370/1975/4517.short}
  {\bibfield  {journal} {\bibinfo  {journal} {Philosophical Transactions of the
  Royal Society A: Mathematical, Physical and Engineering Sciences}\ }\textbf
  {\bibinfo {volume} {370}},\ \bibinfo {pages} {4517} (\bibinfo {year}
  {2012})}\BibitemShut {NoStop}%
\bibitem [{\citenamefont {Ritz}\ \emph {et~al.}(2004)\citenamefont {Ritz},
  \citenamefont {Thalau}, \citenamefont {Phillips}, \citenamefont {Wiltschko},\
  and\ \citenamefont {Wiltschko}}]{ritz2004resonance}%
  \BibitemOpen
  \bibfield  {author} {\bibinfo {author} {\bibfnamefont {T.}~\bibnamefont
  {Ritz}}, \bibinfo {author} {\bibfnamefont {P.}~\bibnamefont {Thalau}},
  \bibinfo {author} {\bibfnamefont {J.~B.}\ \bibnamefont {Phillips}}, \bibinfo
  {author} {\bibfnamefont {R.}~\bibnamefont {Wiltschko}}, \ and\ \bibinfo
  {author} {\bibfnamefont {W.}~\bibnamefont {Wiltschko}},\ }\href
  {http://www.nature.com/nature/journal/v429/n6988/abs/nature02534.html}
  {\bibfield  {journal} {\bibinfo  {journal} {Nature}\ }\textbf {\bibinfo
  {volume} {429}},\ \bibinfo {pages} {177} (\bibinfo {year}
  {2004})}\BibitemShut {NoStop}%
\bibitem [{\citenamefont {Cai}\ \emph {et~al.}(2010)\citenamefont {Cai},
  \citenamefont {Guerreschi},\ and\ \citenamefont {Briegel}}]{cai2010quantum}%
  \BibitemOpen
  \bibfield  {author} {\bibinfo {author} {\bibfnamefont {J.}~\bibnamefont
  {Cai}}, \bibinfo {author} {\bibfnamefont {G.~G.}\ \bibnamefont {Guerreschi}},
  \ and\ \bibinfo {author} {\bibfnamefont {H.~J.}\ \bibnamefont {Briegel}},\
  }\href {http://journals.aps.org/prl/abstract/10.1103/PhysRevLett.104.220502}
  {\bibfield  {journal} {\bibinfo  {journal} {Physical review letters}\
  }\textbf {\bibinfo {volume} {104}},\ \bibinfo {pages} {220502} (\bibinfo
  {year} {2010})}\BibitemShut {NoStop}%
\bibitem [{\citenamefont {Cai}\ and\ \citenamefont
  {Plenio}(2013)}]{cai2013chemical}%
  \BibitemOpen
  \bibfield  {author} {\bibinfo {author} {\bibfnamefont {J.}~\bibnamefont
  {Cai}}\ and\ \bibinfo {author} {\bibfnamefont {M.~B.}\ \bibnamefont
  {Plenio}},\ }\href
  {http://journals.aps.org/prl/abstract/10.1103/PhysRevLett.111.230503}
  {\bibfield  {journal} {\bibinfo  {journal} {Physical Review Letters}\
  }\textbf {\bibinfo {volume} {111}},\ \bibinfo {pages} {230503} (\bibinfo
  {year} {2013})}\BibitemShut {NoStop}%
\bibitem [{\citenamefont {Kominis}(2009)}]{kominis2009quantum}%
  \BibitemOpen
  \bibfield  {author} {\bibinfo {author} {\bibfnamefont {I.~K.}\ \bibnamefont
  {Kominis}},\ }\href
  {http://journals.aps.org/pre/abstract/10.1103/PhysRevE.80.056115} {\bibfield
  {journal} {\bibinfo  {journal} {Physical Review E}\ }\textbf {\bibinfo
  {volume} {80}},\ \bibinfo {pages} {056115} (\bibinfo {year}
  {2009})}\BibitemShut {NoStop}%
\bibitem [{\citenamefont {Gauger}\ \emph {et~al.}(2011)\citenamefont {Gauger},
  \citenamefont {Rieper}, \citenamefont {Morton}, \citenamefont {Benjamin},\
  and\ \citenamefont {Vedral}}]{gauger2011sustained}%
  \BibitemOpen
  \bibfield  {author} {\bibinfo {author} {\bibfnamefont {E.~M.}\ \bibnamefont
  {Gauger}}, \bibinfo {author} {\bibfnamefont {E.}~\bibnamefont {Rieper}},
  \bibinfo {author} {\bibfnamefont {J.~J.}\ \bibnamefont {Morton}}, \bibinfo
  {author} {\bibfnamefont {S.~C.}\ \bibnamefont {Benjamin}}, \ and\ \bibinfo
  {author} {\bibfnamefont {V.}~\bibnamefont {Vedral}},\ }\href
  {http://journals.aps.org/prl/abstract/10.1103/PhysRevLett.106.040503}
  {\bibfield  {journal} {\bibinfo  {journal} {Physical review letters}\
  }\textbf {\bibinfo {volume} {106}},\ \bibinfo {pages} {040503} (\bibinfo
  {year} {2011})}\BibitemShut {NoStop}%
\bibitem [{\citenamefont {Bandyopadhyay}\ \emph {et~al.}(2012)\citenamefont
  {Bandyopadhyay}, \citenamefont {Paterek},\ and\ \citenamefont
  {Kaszlikowski}}]{bandyopadhyay2012quantum}%
  \BibitemOpen
  \bibfield  {author} {\bibinfo {author} {\bibfnamefont {J.~N.}\ \bibnamefont
  {Bandyopadhyay}}, \bibinfo {author} {\bibfnamefont {T.}~\bibnamefont
  {Paterek}}, \ and\ \bibinfo {author} {\bibfnamefont {D.}~\bibnamefont
  {Kaszlikowski}},\ }\href
  {http://journals.aps.org/prl/abstract/10.1103/PhysRevLett.109.110502}
  {\bibfield  {journal} {\bibinfo  {journal} {Physical Review Letters}\
  }\textbf {\bibinfo {volume} {109}},\ \bibinfo {pages} {110502} (\bibinfo
  {year} {2012})}\BibitemShut {NoStop}%
\bibitem [{\citenamefont {Poonia}\ \emph {et~al.}()\citenamefont {Poonia},
  \citenamefont {Saha},\ and\ \citenamefont
  {Ganguly}}]{Poonia2014FunctionalWindow}%
  \BibitemOpen
  \bibfield  {author} {\bibinfo {author} {\bibfnamefont {V.~S.}\ \bibnamefont
  {Poonia}}, \bibinfo {author} {\bibfnamefont {D.}~\bibnamefont {Saha}}, \ and\
  \bibinfo {author} {\bibfnamefont {S.}~\bibnamefont {Ganguly}},\ }\href
  {http://www.ntu.edu.sg/ias/upcomingevents/QuEBS14/Documents/Abstracts%20of%20QuEBS2014-posters.pdf}
  {\bibinfo  {journal} {Workshop on Quantum Effects in Biological Systems,
  Singapore, (2014)}\ }\BibitemShut {NoStop}%
\bibitem [{\citenamefont {Cai}\ \emph {et~al.}(2012)\citenamefont {Cai},
  \citenamefont {Caruso},\ and\ \citenamefont {Plenio}}]{cai2012quantum}%
  \BibitemOpen
\bibfield  {journal} {  }\bibfield  {author} {\bibinfo {author} {\bibfnamefont
  {J.}~\bibnamefont {Cai}}, \bibinfo {author} {\bibfnamefont {F.}~\bibnamefont
  {Caruso}}, \ and\ \bibinfo {author} {\bibfnamefont {M.~B.}\ \bibnamefont
  {Plenio}},\ }\href
  {http://journals.aps.org/pra/abstract/10.1103/PhysRevA.85.040304} {\bibfield
  {journal} {\bibinfo  {journal} {Physical Review A}\ }\textbf {\bibinfo
  {volume} {85}},\ \bibinfo {pages} {040304} (\bibinfo {year}
  {2012})}\BibitemShut {NoStop}%
\bibitem [{\citenamefont {Timmel}\ \emph {et~al.}(1998)\citenamefont {Timmel},
  \citenamefont {Till}, \citenamefont {Brocklehurst}, \citenamefont
  {McLauchlan},\ and\ \citenamefont {Hore}}]{timmel1998effects}%
  \BibitemOpen
  \bibfield  {author} {\bibinfo {author} {\bibfnamefont {C.}~\bibnamefont
  {Timmel}}, \bibinfo {author} {\bibfnamefont {U.}~\bibnamefont {Till}},
  \bibinfo {author} {\bibfnamefont {B.}~\bibnamefont {Brocklehurst}}, \bibinfo
  {author} {\bibfnamefont {K.}~\bibnamefont {McLauchlan}}, \ and\ \bibinfo
  {author} {\bibfnamefont {P.}~\bibnamefont {Hore}},\ }\href
  {http://www.tandfonline.com/doi/abs/10.1080/00268979809483134} {\bibfield
  {journal} {\bibinfo  {journal} {Molecular Physics}\ }\textbf {\bibinfo
  {volume} {95}},\ \bibinfo {pages} {71} (\bibinfo {year} {1998})}\BibitemShut
  {NoStop}%
\bibitem [{\citenamefont {Xu}\ \emph {et~al.}(2014)\citenamefont {Xu},
  \citenamefont {Zou}, \citenamefont {Li}, \citenamefont {Li},\ and\
  \citenamefont {Shao}}]{xu2014effect}%
  \BibitemOpen
  \bibfield  {author} {\bibinfo {author} {\bibfnamefont {B.-M.}\ \bibnamefont
  {Xu}}, \bibinfo {author} {\bibfnamefont {J.}~\bibnamefont {Zou}}, \bibinfo
  {author} {\bibfnamefont {H.}~\bibnamefont {Li}}, \bibinfo {author}
  {\bibfnamefont {J.-G.}\ \bibnamefont {Li}}, \ and\ \bibinfo {author}
  {\bibfnamefont {B.}~\bibnamefont {Shao}},\ }\href
  {http://journals.aps.org/pre/abstract/10.1103/PhysRevE.90.042711} {\bibfield
  {journal} {\bibinfo  {journal} {Physical Review E}\ }\textbf {\bibinfo
  {volume} {90}},\ \bibinfo {pages} {042711} (\bibinfo {year}
  {2014})}\BibitemShut {NoStop}%
\bibitem [{\citenamefont {Baumgratz}\ \emph {et~al.}(2014)\citenamefont
  {Baumgratz}, \citenamefont {Cramer},\ and\ \citenamefont
  {Plenio}}]{baumgratz2014quantifying}%
  \BibitemOpen
  \bibfield  {author} {\bibinfo {author} {\bibfnamefont {T.}~\bibnamefont
  {Baumgratz}}, \bibinfo {author} {\bibfnamefont {M.}~\bibnamefont {Cramer}}, \
  and\ \bibinfo {author} {\bibfnamefont {M.~B.}\ \bibnamefont {Plenio}},\
  }\href {\doibase 10.1103/PhysRevLett.113.140401} {\bibfield  {journal}
  {\bibinfo  {journal} {Phys. Rev. Lett.}\ }\textbf {\bibinfo {volume} {113}},\
  \bibinfo {pages} {140401} (\bibinfo {year} {2014})}\BibitemShut {NoStop}%
\bibitem [{\citenamefont {Doherty}\ \emph {et~al.}(2013)\citenamefont
  {Doherty}, \citenamefont {Manson}, \citenamefont {Delaney}, \citenamefont
  {Jelezko}, \citenamefont {Wrachtrup},\ and\ \citenamefont
  {Hollenberg}}]{doherty2013nitrogen}%
  \BibitemOpen
  \bibfield  {author} {\bibinfo {author} {\bibfnamefont {M.~W.}\ \bibnamefont
  {Doherty}}, \bibinfo {author} {\bibfnamefont {N.~B.}\ \bibnamefont {Manson}},
  \bibinfo {author} {\bibfnamefont {P.}~\bibnamefont {Delaney}}, \bibinfo
  {author} {\bibfnamefont {F.}~\bibnamefont {Jelezko}}, \bibinfo {author}
  {\bibfnamefont {J.}~\bibnamefont {Wrachtrup}}, \ and\ \bibinfo {author}
  {\bibfnamefont {L.~C.}\ \bibnamefont {Hollenberg}},\ }\href
  {http://www.sciencedirect.com/science/article/pii/S0370157313000562}
  {\bibfield  {journal} {\bibinfo  {journal} {Physics Reports}\ }\textbf
  {\bibinfo {volume} {528}},\ \bibinfo {pages} {1} (\bibinfo {year}
  {2013})}\BibitemShut {NoStop}%
\bibitem [{\citenamefont {Maze}\ \emph {et~al.}(2008)\citenamefont {Maze},
  \citenamefont {Stanwix}, \citenamefont {Hodges}, \citenamefont {Hong},
  \citenamefont {Taylor}, \citenamefont {Cappellaro}, \citenamefont {Jiang},
  \citenamefont {Dutt}, \citenamefont {Togan}, \citenamefont {Zibrov} \emph
  {et~al.}}]{maze2008nanoscale}%
  \BibitemOpen
  \bibfield  {author} {\bibinfo {author} {\bibfnamefont {J.}~\bibnamefont
  {Maze}}, \bibinfo {author} {\bibfnamefont {P.}~\bibnamefont {Stanwix}},
  \bibinfo {author} {\bibfnamefont {J.}~\bibnamefont {Hodges}}, \bibinfo
  {author} {\bibfnamefont {S.}~\bibnamefont {Hong}}, \bibinfo {author}
  {\bibfnamefont {J.}~\bibnamefont {Taylor}}, \bibinfo {author} {\bibfnamefont
  {P.}~\bibnamefont {Cappellaro}}, \bibinfo {author} {\bibfnamefont
  {L.}~\bibnamefont {Jiang}}, \bibinfo {author} {\bibfnamefont {M.~G.}\
  \bibnamefont {Dutt}}, \bibinfo {author} {\bibfnamefont {E.}~\bibnamefont
  {Togan}}, \bibinfo {author} {\bibfnamefont {A.}~\bibnamefont {Zibrov}},
  \emph {et~al.},\ }\href
  {http://www.nature.com/nature/journal/v455/n7213/abs/nature07279.html}
  {\bibfield  {journal} {\bibinfo  {journal} {Nature}\ }\textbf {\bibinfo
  {volume} {455}},\ \bibinfo {pages} {644} (\bibinfo {year}
  {2008})}\BibitemShut {NoStop}%
\bibitem [{\citenamefont {Balasubramanian}\ \emph {et~al.}(2009)\citenamefont
  {Balasubramanian}, \citenamefont {Neumann}, \citenamefont {Twitchen},
  \citenamefont {Markham}, \citenamefont {Kolesov}, \citenamefont {Mizuochi},
  \citenamefont {Isoya}, \citenamefont {Achard}, \citenamefont {Beck},
  \citenamefont {Tissler} \emph {et~al.}}]{balasubramanian2009ultralong}%
  \BibitemOpen
  \bibfield  {author} {\bibinfo {author} {\bibfnamefont {G.}~\bibnamefont
  {Balasubramanian}}, \bibinfo {author} {\bibfnamefont {P.}~\bibnamefont
  {Neumann}}, \bibinfo {author} {\bibfnamefont {D.}~\bibnamefont {Twitchen}},
  \bibinfo {author} {\bibfnamefont {M.}~\bibnamefont {Markham}}, \bibinfo
  {author} {\bibfnamefont {R.}~\bibnamefont {Kolesov}}, \bibinfo {author}
  {\bibfnamefont {N.}~\bibnamefont {Mizuochi}}, \bibinfo {author}
  {\bibfnamefont {J.}~\bibnamefont {Isoya}}, \bibinfo {author} {\bibfnamefont
  {J.}~\bibnamefont {Achard}}, \bibinfo {author} {\bibfnamefont
  {J.}~\bibnamefont {Beck}}, \bibinfo {author} {\bibfnamefont {J.}~\bibnamefont
  {Tissler}},  \emph {et~al.},\ }\href
  {http://www.nature.com/nmat/journal/v8/n5/full/nmat2420.html} {\bibfield
  {journal} {\bibinfo  {journal} {Nature materials}\ }\textbf {\bibinfo
  {volume} {8}},\ \bibinfo {pages} {383} (\bibinfo {year} {2009})}\BibitemShut
  {NoStop}%
\end{thebibliography}%
\end{document}